\documentstyle[12pt]{article}

\begin{document}

\begin{titlepage}

\begin{flushright}
TPI-MINN-96/18-T\\
hep-ph/9610236\\
\end{flushright}
\vspace{.3cm}

\begin{center}
{ \Large
{\bf Comments on glueballinos ($R^0$ particles) and $R^0$ searches}
} 
\end{center}  
\vspace*{1cm}  

\begin{center} {\Large Shmuel Nussinov}
\footnote{{\it e-mail:} nussinov@ccsg.tau.ac.il} \\
\vspace*{1.5cm}
{\it Department of Physics, Tel-Aviv University,\\
Tel-Aviv, Israel}\\
\vspace*{.8cm}
and \\
\vspace*{.8cm}
{\it  Theoretical Physics Institute, University of Minnesota,\\
116 Church St. SE, Minneapolis, MN 55455, USA}\\
\vspace{1cm}
\end{center}

\begin{abstract} 

We propose a search strategy for the light $R^0$ (glueballino) particle
suggested by G.Farrar in connection with the light gluino scenario. The basic
idea is to moderate and stop the $R^0$ particles and then observe their decay 
to almost
monochromatic $\pi^0$-s -- at an appropriate time delay relative to a primary
collision event, where a gluino jet, likely to fragment into the $R^0$, was
produced. This technique is optimized at colliders ($\bar pp$, $e^+e^-$), and
depends on qualitative features of the $R^0$ hadronic interactions which we
discuss in detail.
\end{abstract}

\end{titlepage}

\section{Introduction}

Over the last few years a case for light gluino and photino has been made by
G.Farrar \cite{farrar}, \cite{farrar-rutgers}, \cite{cakir}. 
Supersymmetry breaking is expected to generate masses in the 
$100-10^3$ GeV range for the superpartners of the known light fermions and
bosons. (The upper range is a theoretical bias and is natural if SUSY and 
E.W.symmetry breaking are concurrent and related --- the lower limits are
experimental ``bounds''). However in certain schemes the partners of the exactly
massless vector bosons (--associated with unbroken $SU(3)_C\times U(1)_{EM}$
gauge symmetric), namely the photino and gluino, obtain only small $\le$0(GeV)
masses via radiative (loop) correction.

Since finding a light gluino $\tilde g$ at existing,
pre-LHC, accelerators would have such monumental consequences, we believe that
even the remote chance that this scenario is realized in nature is worth
pursuing. 

Light gluinos modify various aspects of perturbative QCD - such as the running
of $\alpha_s(Q^2)$, quarkonium and jet physics \cite{gouvea}. It seems that 
these modifications cannot -- at present -- definitely rule out the light 
gluino hypothesis.

The gluino cannot, because of color confinement, directly manifest as a free
particle, but rather as a constituent of a color neutral
hadron the ``glueballino'' $\tilde gg$. This new hadron, termed $R^{(0)}$
by Farrar, is the lightest hadron with negative $R$ parity. It is therefore
strong interaction stable but decays weakly via squark exchange into a
photino + hadrons
\begin{equation}
R^0\to\tilde\gamma + H^0 
\;(H^0=\pi^0,\;\eta^0,\;\pi^+\pi^-,\;\pi^+\pi^-\pi^0,\;{\rm etc})
\end{equation}
\{If $m_{R^0}\le m_{\tilde\gamma}$ the $R^0$ would be stable altogether. This
theoretically (even more) unlikely scenario is forbidden by the following
consideration: after the $R^0R^0\to$hadrons annihilation freeze out, a relic
density of $R^0$: $\frac{n_{R^0}}{n_\gamma}\simeq10^{-18}$ 
remains. This implies $\frac{n_{R^0}}{n_p}\ge10^{-10}$. As we will argue
below $R^0$ is likely to attach to heavy nuclei. The ($R^0-A,Z$) composites
constitute new exotic ``isotopes'' - which have been excluded with very 
high precision \cite{voloshin}.\}

The $R^0$ decay lifetime scales as the squark mass to the fourth power and is
even more sensitive to the $R^0$ mass (or rather $R^0-\tilde\gamma$ mass
difference). For $R^0$ masses in the range
\begin{equation}
m_{R^0}=1.5 \pm 0.2 \;{\rm GeV}
\end{equation}
and
\begin{equation}
70\;{\rm Gev}\le m_{\tilde q}\le 500\; {\rm GeV}
\end{equation}
Farrar (and we) find a very wide range for $R^0$
lifetimes 
\begin{equation}
\tau_{R^0}\simeq (3\times10^{-11}{\rm sec} - 10^{-4}{\rm sec})
\end{equation}

The $R^0$ hadron is a most striking prediction of the light gluino
scheme. How come such a particle has not been discovered yet?

As correctly pointed out by Farrar \cite{farrar}, \cite{cakir}, searches for 
SUSY particles looking for
missing $E_T$ (transverse missing energy) signals are rather insensitive to
$R^0$-s. Over most of the above lifetime range the $R^0$ interacts in the
calorimeters and loses most of its energy there. The final
decay photino would then carry a tiny ($\le$ few GeV) energy and would be
indistinguishable from a neutrino with similar energy. Likewise the $R^0$ could
not have been discovered in beam dump experiments looking for penetrating
particles. Finally searches for gluino jets based on their $q$ jet like angular
distribution \cite{gouvea} may also be somewhat hampered by the fact that the
leading hadron in these jets, namely $\tilde gg\equiv R^0$, can take most
(60\%-80\%) of the jet's energy. Unlike leading $gg$ glueball in gluon
jets which decay to multi $\pi$-s, $K$-s hadronic, $10^{-24}$ sec, time scales,
the $R^0$ are stable on jet evolution time scale. Gluino jets may therefore be 
less conspicouse than gluon or quark jets.

In principle neutral beam experiments with decay paths at a 
distance $L$ from the production point can
be used to look for $R^0$ decays if $L/c\simeq\gamma_{R^0}\tau_{R^0}$.
Also the $R^0$-s in the beam could be looked for via their hadronic collisions
by careful measurement of time of flight and of $R^0p$ elastic collisions 
kinematics which hopefully can separate the $R^0$ from the dominant neutron 
component \cite{farrar}, \cite{farrar-rutgers}.

We will focus on an alternative approach. The hadronic interactions of the
$R^0$ can be utilized to moderate it. $R^0$ decays delayed by $\tau(R^0)$ from
``relevant'' primary interaction, (chosen on the basis of being likely to send 
an $R^0$ in the detector's direction) could then yield almost monochromatic
$\pi^0$-s, $\eta^0$-s and other $\pi^+\pi^-$, $\pi^+\pi^-\pi^0$ hadronic 
systems.

In the following we will review the basic steps and the features of 
the putative $R^0$ hadrons which control them.

\section{$R^0$ production}
The sensitivity of $R^0$ searches depends on its production rates and on the 
$R^0$ inclusive spectra, both of which we will estimate in the following.

The $\tilde g\tilde g$-gluon coupling is larger than that of 
$\bar qq$-gluon by the ratio of the second Casimir
operators for the octet and triplet representations of $SU(3)_c$:
\begin{equation}
\label{3}
g^2_{\tilde g \tilde g g}=g^2_{ggq}=[C_2({\underline 8})/C_2({\underline 3})]
\;g^2_{\bar qqg}=2.25\;g^2_{\bar qqg}
\end{equation}

This suggests that gluino jets and mini-jets and the $R^0$ particles, that
these jets fragment into, will be more copious than charm jets
and charmed particles. This is even more so when the $R^0$ mass is appreciably
smaller than the $D-D^*$ mass: ${\overline m}_D=\frac{3}{4}m_{D^*}+
\frac{1}{4}m_D=1.97$ GeV. [$m_{R^0}\le {\overline m}_D$ is assumed to always be
the case]. It was found that at the tevatron collider 10\% of all
jets had $D+D^*$ \cite{abe}. We expect at least as many $R^0$-s containing 
jets. 

Unlike the case of $B^{(*)}$ mesons, whose complete production pattern can be
predicted from perturbative QCD, a substantial fraction of $D$-s may be produced
``softly'' with no separate $c{\overline c}$ jets. This is a-fortiori
the case for $R^0$ particle. If this soft component can be described by the 
phenomenological fit
\begin{equation}
\frac{d\sigma}{dp_T}\simeq e^{-6\sqrt{p^2_T+m^2}},
\end{equation}
we would have (for $m_{R^0}\simeq1.5$ GeV) a fairly large ratio
\begin{equation}
\sigma(R^0)/\sigma(D+D^*)=\frac{m_{R^0}e^{-6m_{R^0}}}
{{\overline m}_D e^{-6{\overline m}_D}}\simeq16
\end{equation}
Unfortunately $10-10^3$ more neutrons and neutral kaons ($K_0^L$ specifically)
are produced and provide a strong background for interacting (and decaying)
$R^0$-s, respectively. The $R^0$ particles are produced with larger transverse
momenta than the latter $n$-s, $K^0$-s and the other light 
hadrons. This feature is particularly true for particles produced in
the forward (${\overline p}$) and backward ($p$) fragmentation region and in
the central rapidity plateau (in the ${\overline p}p$ cms). This in turn
causes a broader angular distribution of the produced $R^0$ particles
\begin{equation}
\Theta_R\simeq\frac{p_T(R)}{p(R)}>\frac{p_T(n)}{p(n)}\;[\;{\rm or}\;\;
\frac{p_T(K_L)}{p(K_L)}]\simeq\Theta(n)\;[\;{\rm or}\;\;\Theta(K_L)]
\end{equation}
In particular many of the neutrons and some $K_0$-s may be ``leading''
particles, in the fragmentation of the proton projectile in a fixed target
experiment. These will than have both small $p_T$-s and large
laboratory energies and hence very small angles.

The neutral beams in various fixed target experiments, obtained by strong
collimation in almost exactly the original proton beam direction, 
will therefore be further strongly enriched in neutrons and $K_0^L$. It is thus
particularly difficult to look for the decays or interactions of the relatively
rare putative $R^0$ particles in such beams.

\section{The lifetime of $R^0$}
The most important parameter for determining optimal search strategies for the
$R^0$ particle is its lifetime $\tau_{R^0}$. Indeed as $\tau_{R^0}$ spans the
six--seven decade range indicated in eq.(\ref{3}) its very evolution as it
propagates in matter will drastically change. For $\tau_{R^0}\le10^{-9}$sec
it is likely to decay within a meter or so from the collision point and
manifest via a missing $p_T$ signal. If $\tau_{R^0}\ge10^{-8}$ sec then it is
likely to become non-relativistic and further loose energy by elastic
$R^0$-nuclei collisions before decaying. Let us therefore briefly recall why
these lifetimes are naturally expected (if $m_{R^0}$ is indeed in the
$m_{R^0}=1.5\pm0.2$ GeV range prescribed!).

The gluino decay $\tilde g^0\to\overline qq\tilde\gamma$ with $\bar qq=\bar uu$,
or $\bar dd$, proceeds via the squark exchange diagram.

If the gluino was a physical particle with mass $m_{\tilde g}$ we could use this
Feynman diagram to compute directly the $\tilde g$ decay rate. By comparing
with the $W^-$ exchange in $\mu\to\nu_\mu\nu_ee$ decay we have:
\begin{equation}
\label{7}
\frac{\Gamma(\tilde g)}{\Gamma(\mu)}=
\frac{\alpha_{QCD}(m_{\tilde g})\;\alpha_{em}}{(\alpha_{weak})^2}\cdot
\Bigl(\frac{m_W}{m_{\bar q}}\Bigr)^4\cdot
\frac{5}{3}\cdot
\Bigl(\frac{m_{\bar q}}{m_\mu}\Bigr)^5\;
\frac{r(m^2_{\tilde\gamma}/m^2_{\tilde g})}{r(m^2_e/m^2_\mu)}
\end{equation}
where the above four factors represent coupling constants and propagator
ratios, $N_c(q^2_u+q^2_d)=3.5/9=5/3$ is a charge color factor
summing over $u\bar u,\;\;d\bar d$ states of various charges, and a phase space
ratio. (In particular 
\begin{equation}
r(z)=1-8z+8z^3-z^4-12z^2\ln z
\end{equation}
is relevant in the limit $m_{\tilde\gamma}\gg m^0_u,\; m^0_d$). 

If the gluino had a ``large'', $m^{(0)}_{\tilde g}\ge 1.2$ GeV, bare mass and, 
like the charm quark in $D^+=c\bar d$, dominated the
$R^0=\tilde gg$ mass then, just like in the case of $D^+$ decay, we could use, 
in a
spectator approximation, $m_{\tilde g}\simeq m_{R^0}$ to estimate $\tau(R^0)$.
Eq.(\ref{7}) with $\alpha_{QCD}(1.5\;{\rm GeV})\simeq0.3$,
$\alpha_{em}\simeq0.0073$, $\alpha_W=0.034$ and $m_{\tilde\gamma}\ll m_{R^0}
\simeq m_{\tilde g}$ (so that $(m^2_{\tilde g}/m^2_R)\simeq1)$ yields:
\begin{eqnarray}
\label{9}
\tau(\tilde g)\simeq\tau(R^0)\simeq 1.6\times 10^{-6}\;{\rm sec}\;\;
\Bigl(\frac{m_{\tilde q}}{100\;{\rm GeV}}\Bigr)^4\;
\Bigl(\frac{0.1\;{\rm GeV}}{m_{R^0}(\simeq m_{\tilde g})}\Bigr)^5
\nonumber\\
\simeq2.5\times 10^{-12}\;{\rm sec}\;(m_{\tilde q}/100\;{\rm GeV})^4
\end{eqnarray}
where we used in the last step $m_{\tilde g}\simeq m_{R^0}\simeq 1.5$ GeV. 
However, in the envisioned scenario $m^{(0)}_{\tilde g}\ll m_{R^0}$, and 
eq.(\ref{9})
overestimates the $R^0$ decay rate. Most of the $R^0$ mass is due to QCD
dynamic effects: the chromoelectric (and magnetic) fields i.e. surrounding gluon
cloud and some $\bar qq$ sea. The bare gluino therefore
carries a fraction $xm_{R^0}$ of the $R^0$ mass with a (differential)
probability $f(x)$ normalized to
\begin{equation}
\int\limits_0^1 f(x)\;dx=1
\end{equation}
In a ``generalized spectator'' approximation we assume that the valence gluon
and other gluons and/or $\bar qq$ pairs ``sail along'', as the bare, locally
coupled gluino, of ``mass'' $xm_{R^0}$, decays. This then yields 
\begin{equation}
\label{11}
\tau(R^0)^{-1}=0.4\times 10^{12}({\rm sec})^{-4}\;
\Bigl[\frac{100\;{\rm GeV}}{m_{\tilde g}}\Bigr]^4\cdot
\int\limits_0^1 f(x)\;x^5\;r\left[(y/x)^2\right]\;dx
\end{equation}

The ratio
\begin{equation}
y\equiv\frac{m_{\tilde\gamma}}{m_{R^0}}
\end{equation}
controls further phase-space suppression doe to finite photino mass via
$r\left[(y/x)^2\right]=r\left[\frac{m^2_{\tilde\gamma}}{(xm_{R^0})^2}\right]$.
It is convenient to parameterize the decreasing probability
that the gluino carries most of the total mass by power-like vanishing of
$f(x)$ as $x\to1$
\begin{equation}
f(x)\simeq(1-x)^\gamma(\gamma+1)
\end{equation}

Even for $m_{\tilde\gamma}=0$ the
last  factor in eq.(\ref{11}) above supplies sizable suppression: 

\begin{equation}
I_\gamma(0)=\int\;f(x)\;x^5\;dx=\frac{(\gamma+1)!\;5!}{(\gamma+5+1)!}
\end{equation}
$I_\gamma(0)$ is smaller than $\sim0.01$ for $\gamma\ge3$. Clearly 
$I_\gamma(y)\equiv\int(1-x)^\gamma\;x^5\;r\left[(y/x)^2\right]<I_\gamma(0)$
could be arbitrary small (or zero!) once $m^0_\gamma\to m_{R^0}$ i.e. $y\to1$.
Following Farrar we will adopt the range of lifetimes:
\begin{equation}
\label{16}
\tau_{R^0}=(10^{-10}-10^{-7})\;(m_{\tilde g}/100\;{\rm GeV})^4 \;{\rm sec}
\end{equation}
corresponding to $2\times10^{-5}\le I\le2\times10^{-2}$.

In the light gluino scenario the lower experimental bounds on squark masses,
which are based on missing energy searches, do not apply.
The squark decays (in $\simeq10^{-26}$ sec) to a quark and gluino. The gluino,
after forming the $R^0$, deposits most of its energy. 

For $\tau(R^0)$ estimates
we will therefore take squark masses in the range
\begin{equation}
60\;{\rm GeV}\le m_{\tilde g}\le500\;{\rm GeV}
\end{equation}

The lower limit is based on the modification of $R\equiv \sigma_{e^+e^-\to
{\rm hadrons}}/\sigma_{e^+e^-\to\mu^+\mu^-}$ measured at LEP II due to 
$\tilde q \bar{\tilde q}$ production. The upper bound is a theoretical bias.
Substituting in eq.(\ref{16}) we find
that the wide lifetime range $3\times10^{-11}\;{\rm sec}\le\tau_{R^0}\le10^{-4}$
sec is indeed theoretically expected. 

The pattern of photino couplings implies
that the $\bar qq$ state in the $\tilde g\to\bar qq+\tilde\gamma$ decay is
\begin{equation}
q_u\bar uu+q_d\bar dd=\frac{2}{3}\bar uu-\frac{1}{3}\bar dd
\end{equation}
or, if we include the s quark,
\begin{equation}
q_u\bar uu+q_d\bar dd\simeq\frac{2}{3}\bar uu-\frac{1}{3}\bar dd
-\frac{1}{3}\bar ss
\end{equation}
This state overlaps more strongly with the $\pi^0$ state
$$\pi^0=(\bar uu-\bar dd)/\sqrt{2}$$
than with $\eta^0$
$$\eta^0\simeq(\bar uu+\bar dd-\bar ss)/\sqrt{2}$$
In the absence of any dynamical ``form factor'', effects this would enhance the
$R^0\to\pi^0\tilde\gamma$ over $R^0\to\eta^0\tilde\gamma$ by at least a factor
of nine, i.e.
\begin{equation}
\Gamma(R^0\to r^0\tilde\gamma)/\Gamma(\eta^0\to r^0\tilde\gamma)\simeq 9
\end{equation}
However, if $m_{R^0}-m_{\tilde\gamma}\simeq1.4\pm0.3$ is on the high site of
its range, it would appear that higher multiplicity final states
$R^0\to\pi^+\pi^-\tilde\gamma,\;\;\pi^+\pi^-\pi^0\tilde\gamma$ would be
preferred over both $R^0\to\pi^0,\eta^0,+\tilde\gamma$. A crude estimate for the
probability of the two body decay modes can be inferred from
\begin{equation}
Br(D^+\to K{\pi\atop\rho})\simeq{{3\%}\atop{7\%}}
\simeq Br(R^0\to\tilde\gamma\pi^0)
\end{equation}
We will assume that
\begin{equation}
Br(R^0\to\tilde\gamma{{\pi^0}\atop{\eta^0}})=0.1\pm 0.05
\end{equation}

\section{Hadronic $R^0$ interactions}
An $R^0$ hadron propagating in matter scatters from the nuclei. These
interactions are elastic if the kinetic energy of the $R^0$ is below the
binding energy of nucleons
\begin{equation}
\label{22}
T_{R^0}\le B.E.\le10 \;{\rm MeV}
\end{equation}
and quasi-elastic [i.e. nucleons knock out] if:
\begin{equation}
\label{23}
10\;{\rm MeV} \le T_{R^0}\le 0.5\;{\rm GeV}
\end{equation}
Finally at higher energies
\begin{equation}
\label{24}
T_{R^0}>0.5\;{\rm GeV}
\end{equation}
we have, in addition to the (quasi-) elastic $R^0N$scatterings, also genuinely
inelastic scattering involving production of pions, kaons etc. Most of these
interactions are soft -- at least in the $t$ channel -- and cannot be reliably
calculated via perturbative QCD. Yet a qualitative understanding of these
interactions and, in particular, the degree to which these differ from pion,
kaon and nucleon interactions is required, if we wish to estimate the evolution
of energetic (say $E_{R^0}\simeq P_{R^0}\simeq20$ GeV) $R^0$-s in matter.

At very high energies it is believed that meson and nucleon scattering are
dominated by the exchange of (two or more) gluons. At lower (Laboratory)
energies $E_{\rm Lab}\le10$ GeV quark anti-quark annihilations and exchange play a
dominant role. In pre-QCD nomenclature these are the ``Pomeron'' and
mesonic ``Regge trajectories'' exchanges respectively with corresponding
cross section:
\begin{equation}
\sigma_{\rm pom\simeq glue-ex}\simeq{\rm const}
\;(\;{\rm or}\;\;\ln W\;), \qquad W\to\infty
\end{equation}
\begin{equation}
\sigma_{\rm Regge\simeq quark-ex}\simeq\frac{\rm const'}{W}, \qquad W\to\infty
\end{equation}

What are the corresponding contributions for the case of $R^0N$ scattering? The
fact that gluino (and gluon) have stronger gluonic couplings than quarks
(eq.(\ref{3}) suggests that the gluonic exchange (pomeron) part is enhanced
there. Indeed the contribution of the chromomagnetic-electric 
cloud energy 
\begin{equation}
I_{R^0}=\int\;d^3r\;(E^2+B^2)\;[{\rm In}\;R^0\;{\rm state}]
-\int\;[{\rm In}\;{\rm vac}]\;\simeq m_{R^0}-m^0_{\tilde g}
\end{equation}
makes up most of the $R^0$ mass. 
$I_{R^0}$ is larger than the corresponding integral in a meson or even in a 
nucleon. This conforms to the enhanced contributions of gluonic cloud 
interaction to $R^0-N$ scattering.

On the other hand we have no analog of the quark annihilation or exchange
diagrams for $R^0-N$ scattering. [We could have only gluon exchange
analogies of these diagrams and even these only for say $R^0R^0$ scattering].

Most of $M-N$ and $NN$ scatterings for $E_{\rm Lab}\le10$ GeV is
not only merely accounted for by Reggeon exchange. There are further detailed
hadronic models involving pion and other meson
exchanges for the specific reaction mechanism in various inelastic channels.
Since to lowest order there are no such pion (meson) exchanges contributing to
$R^0-N$ scattering we expect the latter to be smaller, particularly for
$E_{lab}(R^0)\le10$ GeV. Further arguments suggest that 
\begin{equation}
\sigma(R^0N)\le \sigma(MN)
\end{equation}
holds not only for $E_{lab}(R^0)\le10$ GeV, but also at higher energies. The
issue of high energy $MN-NN$ cross sections is not completely settled. It is
possible that ``soft pomeron'' contributions due to multiperipheral pion
exchange type models make up a sizable part of the total $M-N$ cross section
even at lab energies as high as 20-40 GeV.

A closely related issue is the ambiguity of which meson nucleon scattering
cross section should we compare $\sigma(R^0N)$ with. These $MN$
cross sections depend on whether we use
\begin{equation}
\label{28-a}
M=\pi=u\bar d \quad:\quad \sigma_{tot}(\pi^+)\simeq 30\;{\rm mb},
\end{equation}

\begin{equation}
\label{28-b}
M=K=u\bar s \quad:\quad 
\frac{1}{2}[\sigma_{tot}(K^+p)+\sigma_{tot}(K^-p)]\simeq 20\;{\rm mb},
\end{equation}
or
\begin{equation}
\label{28-c}
M=\phi=s\bar s \quad:\quad  \sigma_{tot}(\phi)=10\;{\rm mb}.
\end{equation}

\{The total $\phi N$ cross section in eq.(\ref{28-c}) is inferred from vector
dominance in high energy $\phi$ photo-production. Estimates of 
$\sigma_t(\rho p)$ or $\sigma_t(wp)$ done in a similar way for $\rho$, $w$
photoproduction yield $\sigma(\rho^0N)\sim\sigma(w^0N)\simeq\sigma(\pi^\pm N)$
as expected for the $\rho$ and $w$ which like the pion $\pi$ are made of light 
($\bar uu\pm\bar dd$) quarks.\}

The clear cut systematics: 
\begin{equation}
\label{29}
\sigma_{\pi N}>\sigma_{KN}>\sigma_{\phi N}
\end{equation}    
is somewhat puzzling from the point of view of the gluon exchange model as the
QCD coupling are flavor independent and $g^2_{\bar uug}=g^2_{\bar ddg}=g^2_{\bar
ssg}$. One may try and argue that the slightly larger
constituent mass of the $s$ quark ($\sim500$ MeV) as compared with $\sim350$ MeV
for the $u,d$ quarks, can cause -- via reduced mass effects -- the sizes and
cross sections to differ. Even if correct, this explanation undermines our 
original argument
for an enhanced gluon exchange contribution to $R^0N$ asymptotic cross section
in the first place. The ``constituent gluon'' mass -- if this is indeed a
useful concept -- and a-fortiori that of the gluino, are presumably
significantly higher [say: $m_g,\;m_{\tilde g}=0.7,\;0.9$ GeV], than those of
the $s$ quark. The naive modeling of non-relativistic $\tilde g-g$ states would
suggest that these larger constituent masses (and stronger attractive forces)
make a much smaller state. The smaller length of the ``color dipole'' would
then compensate for the $\sim\sqrt{9/4}=3/2$ larger color charges. The 
systematics (\ref{29}) could reflect mainly the weaker contributions of pion
exchange in $KN$ and (even more!) in $\phi N$ scattering, suggesting reduced 
$R^0N$ cross sections as well.

One final piece of evidence for small $R^0$-nucleon cross section comes from
recent HERA experiment where a small ``pomeron-nucleon'' total cross section
has been inferred. This could be relevant to our discussion if indeed the
pomeron is related to two gluon states.

In the ensuing discussion of $R^0$ evolution in matter we will still allow a
wide range of variation for the inelastic $R^0-N$ cross section
\begin{equation}
40\;{\rm mb}\;\sim1.4\sigma^I(NN)\ge\sigma^I(R^0N)\ge0.7\sigma(\phi N)
\simeq7\;{\rm mb} \qquad {\rm For}\;T_{R^0}\ge0.5\;{\rm GeV}
\end{equation}    
and similarly for the elastic cross section
\begin{equation}
15\;{\rm mb}\;\ge\sigma^{El}(R^0N)\ge3\;{\rm mb} \qquad 
{\rm For}\;T_{R^0}\ge0.5\;{\rm GeV}
\end{equation}    
We should keep in mind, however, that the lower values are theoretically more
favorable. Another crucial parameter controlling the hadronic shower initiated
by an energetic $R^0$ is the fraction of the initial $R^0$ momentum, 
$\bar x_F(R^0)$, carried
on average by the final $R^0$ in an inelastic reaction:
\begin{equation}
R^0({\rm momentum}\;p\gg m_{R^0})+N({\rm at}\;{\rm rest})\to N'+n\;{\rm pions}
+R^0(p_f)
\end{equation}    
In proton-proton collisions a final ``leading'' nucleon
(i.e. proton or neutron) carries on average $\bar x_F(p)\simeq1/2$. This
presumably reflects the tendency of a ``diquark'' in the initial nucleon to
reappear as a constituent diquark in the final nucleon.

A key to the evolution of $R^0$ in matter is that the initial gluino in $R^0$
($=\tilde gg$) must be conserved in all collisions. This implies that in any
reaction initiated by an $R^0$ particle there must be an outgoing $R^0$
particle containing the initial $\tilde g$. Since the $R^0$ is likely to
contain also the ``initial constituent gluon'', this leading $R^0$ may carry
even a larger fraction of the incoming momentum than an outgoing nucleon
\begin{equation}
\label{33}
\bar x_F(R^0)\simeq0.65\pm0.15
\end{equation}    
Indeed a large $\bar x_F$ is also consistent with a ``hard fragmentation'' of
gluino jets into a ``strongly leading'' $R^0$ carrying a large fraction of the
gluino jet momentum. Such a hard $\tilde g\to R^0$ fragmentation may be
required in order to avoid the exclusion of the light gluino hypothesis by the
analysis of 4 jet LEP events (see ref.\cite{gouvea}).

At intermediate energies [eq.(\ref{23})] the $R^0$ ``knocks out'' a nucleon out 
of the nucleus. The fraction of kinetic energy retained by $R^0$ can be
approximately obtained just from the two body elastic $R^0N$ collision
kinematics:
\begin{equation}
p_1(R^0)+p_2(N)\to p'_1(R^0)+p'_2(N)
\end{equation}    
The kinetic energy of the final nucleon $T'_2=E'_2-m_N$ is given
in terms of the invariant momentum transfer $t=q^2$, $q=(p_1-p'_1)$ via
\begin{equation}
T'_2=t/2m_N
\end{equation}    
In terms of the $R^0-N$ c.m.s. momentum $p^*$ and scattering angle $\theta^*$,
$t$ is given by 
\begin{equation}
t=2(p^*)^2(1-\cos\theta^*)
\end{equation}    
Expressing $p^*$ in terms of the initial kinetic energy of the incoming $R^0$,
we finally find (for $(p^*/m_N)^2<1$) that:
\begin{equation}
T'_2=T_1-T'_1=\frac{2m_Nm_{R^0}}{(m_N+m_{R^0})^2}\;(1-\cos\theta^*)T_1
\end{equation}    
The retained energy fraction $T'_1/T_1$ is thus:
\begin{equation}
\label{38}
T'_1/T_1=1-\frac{2m_Nm_{R^0}}{(m_N+m_{R^0})^2}\;(1-\cos\theta^*)
\end{equation}    
so that on average we have for $m_{R^0}=1.5$ GeV:
\begin{equation}
\langle T'_1/T_1\rangle_{R^0-N}=1-\frac{2m_Nm_{R^0}}{(m_N+m_{R^0})^2}\;
(1-\overline{\cos\theta^*})\simeq1-0.48\;(1-\overline{\cos\theta^*})
\end{equation}    
If the c.m.s. angular distribution is isotropic then
$\overline{\cos\theta^*}=0$. This indeed will be the case at the lower end of
the range considered where the scattering is mainly in $S$ wave. If we allow a
small admixture of $P$ waves then
\begin{equation}
\overline{\cos\theta^*}=2(\sin\delta_1/\sin\delta_0),
\end{equation}    
where $\delta_0$, $\delta_1$ are the $S$, $P$ wave phase shifts. Assuming
\begin{equation}
0\le\sin\delta_1/\sin\delta_0\le0.2
\end{equation}
we will have
\begin{equation}
\label{43}
\langle T'_1/T_1\rangle=0.62\pm0.1
\end{equation}    
The ratio of the corresponding momenta or velocities is then
\begin{equation}
p'_1/p_1\simeq\beta'_1/\beta_1\simeq\sqrt{T'_1/T_1}\simeq0.79\pm0.16.
\end{equation}    

Finally we address the region of slow ($T_{R^0}\le10$ MeV) $R^0$ particles,
which should be discussed in terms of $R^0$-nucleus (rather than $R^0$-nucleon)
interactions.

The evolution of $R^0$ particles in this stage critically depends on the
existence of a bound ($R^0-A,Z$) state of the $R^0$ and various (Fe, Ca,
S, Si and O) nuclei and on the probability that such bound states in reactions 
like
\begin{equation}
\label{45}
R^0+(A,Z)\to\left\{ {[R^0(A-1,Z)]+n}\atop{[R^0(A-1,Z-1)]+p} \right\}.
\end{equation}    
For the long wavelength slow $R^0$ the individual nucleons will
effectively merge and we expect that its interaction with the nucleus can be
described via a smooth, effective, potential:
\begin{equation}
V_{R^0}(r)=V_\infty\;\frac{\rho(r)}{\rho(\infty)}
\equiv V_0\;\frac{\rho(r)}{\rho(\infty)}
\end{equation}    
where $\rho(r)$ is the nuclear density and $\rho(\infty)$, $V_\infty(=V_0)$ 
refer to an ideal case of infinite uniform nuclear matter.

To simplify our estimates we will approximate also finite nuclei by a
``square-well'' potential:
\begin{equation}
V_{R^0-(A,Z)}(r)=V_0\cdot\Theta\left(R(A,Z)-r\right)
\end{equation}    
with $r$ the distance of $R^0$ from the center of the nucleus and $R(A,Z)$ 
the nuclear radius:
\begin{equation}
R(A,Z)\simeq1.2\;A^{1/3}\;{\rm Fermi}
\end{equation}
If the momentum of the incoming $R^0$
\begin{equation}
p=\sqrt{2m_{R^0}T_{R^0}}
\end{equation}      
is sufficiently low so that
\begin{equation}
\label{50}
p\;R(A,Z)\le1-2,
\end{equation}  
we may assume that the $R^0$ nuclear scattering is dominated by $S$ wave. {The
last condition is satisfied for the nuclei considered with $56\ge A\ge16$ only
if $T_{R^0}\le10$ MeV, which is indeed assumed here. Also eq.(\ref{50})
justifies the smearing of individual potentials of $R^0$ interaction with the
various nucleons}. 

The $S$ wave phase shift is given by 
\begin{equation}
\label{51}
\delta_0=-p\;R+\arctan\Bigl\{\frac{p}{\sqrt{p^2+2mV_0}}
\cdot\tan(\sqrt{p^2+2mV_0\cdot R})\Bigr\}
\end{equation}      
The single parameter $V_0=V_{R^0}$ summarizes our ignorance of $R^0$-nuclear
interactions. We make the reasonable assumption that
\begin{equation}
\label{52}
V_{R^0}\le|V_\Lambda|\simeq50\;{\rm MeV}
\end{equation}      
where $V_\Lambda$ is the attractive potential seen by a $\Lambda$ particle in a
hypernucleus. Indeed the $R^0$, just like the $\Lambda$ particle, is not
affected by the Pauli principle and can migrate towards the inner shells.
However, the $R^0-N$ interactions -- lacking meson (and in particular pion) 
exchanges -- are expected to be weaker. 

Indeed $\pi$ exchange diagrams contribute via 3 body interactions, to the 
$\Lambda$-nuclear binding. We have no analog of this for the $R^0$ case.
As we will argue below (towards the end of this section), the ``$R^+$'' 
companion of the $R^0$ particle is likely to be heavy. In particular
\begin{equation}
\label{53}
m_{R^+}-m_{R^0}\ge m_\pi
\end{equation}      
as compared with
\begin{equation}
\label{54}
m_\Sigma-m_{\Lambda^0}\simeq70\;{\rm MeV}
\end{equation}      
suppressing the three body putative diagrams and $V_{R^0}$. 

The following values of
$V_{R^0}$, consistent with the bound $|V_{R^0}|\le 50$ MeV, lead to
qualitatively different behaviors of $R^0$-nuclear scattering in the
$T_{R^0}\le10$ MeV range: 
\begin{description}
\item[(i)]
$V_{R^0}$ is positive or negative, but very small:
\begin{equation}
|V_{R^0}|\ll T_{R^0}\simeq10\;{\rm MeV}
\end{equation}

\item[(ii)]
$V_{R^0}$ is positive (repulsive) and 
\begin{equation}
V_{R^0}\simeq T_{R^0}\simeq10\;{\rm MeV}
\end{equation}

\item[(iii$)^*$]
$V_{R^0}$ is negative (attractive) and of magnitude
\begin{equation}
2-4\;{\rm MeV}\le|V_{R^0}|\le10\;{\rm MeV}
\end{equation}

\item[(iv$)^*$]
$V_{R^0}$ is negative and 
\begin{equation}
50\;{\rm MeV}>|V_{R^0}|>10\;{\rm MeV}
\end{equation}
 
\item[(v)]
$V_{R^0}$ is positive and 
\begin{equation}
50\;{\rm MeV}>|V_{R^0}|\le10\;{\rm MeV}
\end{equation}
\end{description}
We will next discuss each of these cases.

\begin{description}
\item[(i)]
Expanding the expression (\ref{51}) for $\delta_0(p)$ in the small quantities
$mV_0R^2$ and $mV_0/p^2$ we find
\begin{equation}
\delta_0^{(i)}\simeq-\frac{mV_0}{p^2}(\tan pR-\frac{pR}{\cos^2pR})
\end{equation}  
Thus for either sign of $V_0$ we obtain fairly small $R^0$ nuclear cross section
\begin{equation}
\sigma^{i}(R^0-{\rm Nucleus})\simeq\frac{4\pi(mV_0)^2}{p^6},
\qquad \frac{d\sigma}{d\Omega}={\rm const}
\end{equation}     

\item[(ii)]
In this case $\delta_0$ can be appreciable 
\begin{equation}
\sigma^{(ii)}(R^0-N)\simeq\frac{4\pi}{p^2}\simeq\pi(R_{A,Z})^2,
\qquad \frac{d\sigma}{d\Omega}={\rm const}
\end{equation}     

\item[(iii$)^*$]
The starred cases (iii$)^*$ (and (iv$)^*$) are distinguished by two
important features. First we are guaranteed to have one (or more) bound states.
Indeed the condition for having at least one $S$ wave bound state:
\begin{equation}
\sqrt{2mV_0}\;R(A,Z)\ge\pi/2
\end{equation}   
amounts to
\begin{equation}
V_{R^0}\ge4-2\;{\rm MeV}
\end{equation}     
for nuclei in the oxygen-iron range i.e. for
\begin{equation}
16\le A\le56
\end{equation}     
Second we will argue later that cases (iii) or (iv) are the most likely in any
event. 

For case (iii) we expect enhanced $S$ wave scattering cross section, i.e.
\begin{equation}
\sigma^{(iii)}\simeq\sigma^{(iv)}\simeq\frac{4\pi}{p^2}
\end{equation}  
However the small binding energy ($|\epsilon|\le|V_0|$) may kinematically 
disallow
(or suppress) the actual $R^0$ capture reaction (\ref{45}) since there may
not be sufficient energy for knocking a nucleon out. The latter requires in
particular that 
\begin{equation}
\label{67}
T_{R^0}+|\epsilon|\ge\;{\rm B.E.}\;{\rm of}\;{\rm nucleon}\sim8\;{\rm MeV}
\end{equation}     

\item[(iv$)^*$]
In this case we could have several $S$ wave (and perhaps even $P$ wave) bound
states. However the most important point is that eq.(\ref{67}) is now satisfied,
sometimes with a wide ``margin''. 

The $R^0$ capture -- reaction (\ref{45}), leading to the
formation of the $R^0-(A-1)$ bound state, becomes then
exothermic and the capture cross section is enhanced by the ratio of outgoing 
nucleon momentum and the incoming $R^0$ momentum:
\begin{equation}
\label{68}
\frac{p_N(final)}{p_{R^0}}\simeq\sqrt{
\frac{T_{R^0}+|\epsilon|-{\rm B.E.}}{T_{R^0}}
\cdot\frac{m_n}{m_{R^0}}}\ge1
\end{equation}     
The total scattering cross section in this case $\sigma_{tot}^{(iv)}$ is larger
than $\sigma_{tot}^{(v)}$. The last case corresponds to a flipped (repulsive) 
potential:
$V^{(v)}=-V^{(iv)}$. Hence (see sec.(v) next)
\begin{equation}
\sigma_{tot}^{(iv)}\ge\sigma_{el}^{(v)}\simeq\pi R^2_{(A,Z)}
\end{equation}     
The kinematic enhancement of eq.(\ref{68}) suggests that the capture component 
in the total cross section 
\begin{equation}
\sigma_{tot}^{(iv)}=\sigma_{el}^{(iv)}+\sigma_{cap}^{(iv)}
\end{equation}     
is appreciable.

\item[(v)]
In this case the nucleus acts like a hard sphere. Furthermore
$pR(A,Z)>1$, and we may use the classical cross section
\begin{equation}
\sigma^{(v)}=\pi R^2_{(A,Z)}.
\end{equation}     
\end{description}

For elastic $R^0$-nucleus collision the energy fraction carried by the outgoing
$R^0$ is given by eq.(\ref{38}), but with $m_N\to m(A,Z)\simeq Am_N$. Thus the
average kinetic energy fraction retained
\begin{equation}
\label{72}
\langle T'_1/T_1\rangle_{R^0-{\rm Nucleus}}\simeq1-\frac{2uA}{A^2+2uA+u^2}
\cdot(1-\overline{\cos\theta^*})
\end{equation}     
with
\begin{equation}
u=\frac{m_{R^0}}{m_N}=1.6\pm0.3
\end{equation}     
is significantly closer to one that in the case of elastic $R^0$-nucleon (a
feature well known from neutron moderation). Eq.(\ref{72}) implies a
corresponding ratio for outgoing and incoming $R^0$ velocities:
\begin{equation}
\label{74}
\langle\beta'_1/\beta_1\rangle_{R^0-{\rm Nucleus}}=1-\frac{uA}{A^2+2uA+u^2}
\cdot(1-\overline{\cos\theta^*})
\end{equation}     

We next proceed to estimates of $V_0$ -- making first an argument for negative
i.e. attractive $V_0$. Approximating $R^0\simeq\tilde g+$gluons, i.e.
neglecting additional $\bar qq$ constituents, there are no Pauli exclusion
repulsive effects in $R^0N$ scattering. In fact $\bar qq$
annihilations and quark exchanges i.e. ordinary mesonic contributions to
$R^0-N$ scattering, repulsive or attractive, are altogether absent. The only
interactions are gluonic exchanges i.e. the QCD analogue of ``Van-der-Waals
type'' interactions. Color confinement implies a finite lowest mass
$m^{(0)}_{``2g''}>0$ in the gluon exchange channel. This will modify the 
$r^{-6}-r^{-7}$ potentials derived in the perturbative two massless gluon
exchange approximation by an exponential factor
\begin{equation}
e^{-m^0_{2g}r}
\end{equation}
We conjecture however that the generic property of the $2\gamma$, $2g$
perturbative exchange potentials, namely their almost universal attractive
character, will survive in the full-fledged QCD treatment of
$R^0N$ interactions.

In the perturbative approximation the ``Casimir-Polder''
interaction $V^{(0)}_{2g}$ between two hadrons A and B is
proportional to the following combination of electric and magnetic
polarizabilities of A and B \cite{feinberg}:
\begin{equation}                   
C_{AB}=-\Bigl[\alpha_E^{(A)}\;\alpha_E^{(B)}+\alpha_M^{(A)}\;\alpha_M^{(B)}
-(\alpha_E^{(A)}\;\alpha_M^{(B)}+\alpha_M^{(A)}\;\alpha_E^{(B)})\Bigr]
\end{equation}
For A and B hadrons which are ground states in the respective channels, second
order perturbation in an $E(B)$ background fields, say
$\Delta_E^{(i)}\simeq\alpha_E^{(i)}E^2$, imply negative $\alpha_E^i$ (or
$\alpha_M^i$). Hence $C_{AB}$ is guaranteed to be negative once at least one of
the following holds:
\begin{description}
\item[(i)]
Chromoelectric effects dominate: $\alpha_E^{(i)}\gg\alpha_M^{(i)}$

\item[(ii)]
$\alpha_E^{(A)}=\alpha_E^{(B)}$ and $\alpha_M^{(A)}=\alpha_M^{(B)}$, or at
least 
\begin{equation}
\label{77}
\alpha_E^{(A)}/\alpha_M^{(A)}=\alpha_E^{(B)}/\alpha_M^{(B)}
\end{equation}
\end{description}
Hadronic constituents are generally more relativistic than valence atomic
electrons so that chromo-magnetic effects are more important. Also three and
more gluon exchange are more important relative to two photon exchange. Thus
the attractive character of the $R^0-N$ gluonic interactions is not obvious. 
We proved \cite{niss} via rigorous ``QCD Inequalities'' techniques that the low
energy interaction between ground state pseudoscalars
\begin{equation}
\label{78}
M_{ab}=``\bar q_a\gamma_5 q_b``,\qquad M'_{cd}=``\bar q_c\gamma_5 q_d``
\end{equation}     
is always attractive if
\begin{equation}
m_a=m_c\quad{\rm and}\quad m_b=m_d
\end{equation}     
but all flavors $a,b,c,d$ are different. The latter condition does indeed
exclude any $q$ exchange or $\bar qq$ annihilation so that this $MM'$
attraction refers specifically to the gluonic interactions of interest.

We believe that the same result holds for the interaction between any two
hadrons of similar color structure - reminiscent indeed of condition (\ref{77})
for the perturbative case. 

The $\tilde gg=R^0$ and $N=uud$ nucleon states are clearly different. There is,
however, no obvious reason why the sign of their mutual gluonic interactions
be reversed relative to the gluonic interactions for $NN$ or $R^0R^0$. We will
therefore assume a negative $V_0$. Unfortunately the magnitude $|V_0|$ is not
known. We have suggested above (eq.(\ref{52})) that
$|V_0^{(R)}|\le|V_0^{(\Lambda)}|\simeq40-50$ MeV. We will argue next for the
following likely lower bound: 
\begin{equation}
\label{80}
|V_0^{(R^0)}|\ge10\;{\rm MeV}
\end{equation}     
We express the fact that most of the $R^{(0)}$ mass originates from QCD dynamics in
the following suggestive manner. 
\begin{equation}
\label{81}
1\;{\rm GeV}\le m_{R^0}-m^{(0)}_{\tilde g}\simeq\int\limits_{{\rm
``over}\;R^0"}\;d^3\vec r\;(\vec E_c^2+\vec B_c^2)
\end{equation}     
In the presence of nuclear matter the last expression will be modified. This
modification can be parameterized by the chromo-dielectric (magnetic) constants
shifting away from unity. Assuming that $\int\vec E^2$ dominates, we have for 
the rest energy shift, i.e. for $V_0$, the following expression:
\begin{eqnarray}
\label{82}
V_0=(\frac{1}{\epsilon}-1)\;\int\vec E^2\;d^2\vec r
=(\frac{1}{\epsilon}-1)\;(m_{R^0}-m^0_{\tilde g})
\nonumber\\
\simeq-(\epsilon-1)(m_{R^0}-m^0_{\tilde g})\le-(\epsilon-1)\;{\rm GeV}
\end{eqnarray} 
where we used eq.(\ref{81}) and in the last step expanded in
\begin{equation}
\epsilon-1\ll1
\end{equation}
We note in passing that $V_0<0$ amounts to the natural assumption that
\begin{equation}
\epsilon-1\ge0
\end{equation}     
To estimate $\epsilon-1$ we use the analogue of the familiar QED expression
\begin{equation}
(\epsilon-1)=\frac{4\pi}{3}\;n\;\alpha_E^{(N)}
\end{equation}     
with $\alpha_E^{(N)}$ the (chromo-) electric polarizability of the nucleon and
\begin{equation}
n=(1.2\;{\rm Fermi})^{-3}
\end{equation}    
the nucleon number density. Ordinary electric polarizability can be written as
\begin{equation}
\label{87}
\alpha_E\simeq r_0^3
\end{equation}     
where in the last expression $r_0$ is an effective size [eq.(\ref{87}) also
represents the classic polarizability of a conducting sphere of radius $r_0$]. 
Collecting all the above results we find
\begin{equation}
|V_0|\ge2\Bigl(\frac{r_0^{N,Eff}}{\rm Fermi}\Bigr)^3\;{\rm GeV}
\end{equation}     
Thus $|V_0|\ge10$ MeV once the mild requirement
\begin{equation}
r_0^{N,Eff}\ge0.17\;{\rm Fermi}
\end{equation}     
is satisfied.

Before concluding this section we would like to reiterate the feature which is
the most important distinction between the $R^0$-Nucleon scattering and the
scattering of any other stable hadron on a nucleon.

All the hadrons with lifetimes $\ge10^{-10}$ sec (with the exception of
$\Lambda^0$) come in isospin multiplets:

$$\left(\begin{array}{c}\pi^+ \\ \pi^0 \\ \pi^- \end{array}\right)\quad
{K^+\choose K^0}\quad {{\bar K^0}\choose K^-}\quad {p\choose n}\quad 
\left(\begin{array}{c}\Sigma^+ \\ \Sigma^0 \\ \Sigma^- \end{array}\right)\quad
\left(\begin{array}{c}\Xi^+ \\ \Xi^0 \\ \Xi^- \end{array}\right)$$
Also the $\Lambda^0$ is associated in $SU(3)$ flavor, with the
$\Sigma$ multiplet from which it is split by only 70 MeV (see eq.(\ref{54})).

In principle we could have in addition to $R^{(0)}=\tilde gg$ other more
complex hadrons containing a gluino which could be charged like
\begin{equation}
R^+=\tilde g(u\bar d)_8
\end{equation}     
where as indicated the $\bar ud$ are in a color octet combination. However the
$R^+$ is definitely not an I-spin partner of $R^0$, but rather the lowest
member of a completely different family of particles. These particles are the
SUSY partners of the ``Meikton'' or ``Hemaphrodite'' hadronic states,
which have been speculated by various authors, namely 
$g(u\bar d)_8$, just as $R^0=\tilde gg$ is the SUSY partner of the glueball
state: $gg$.

To date we have no clear-cut evidence for either glue-ball or
``Meikton-Hemaphrodite'' states. Theoretical prejudice suggests
that the glue-balls are significantly lighter:
\begin{equation}
m^0(gg)<m^0\left[g(u\bar d)_8\right]
\end{equation}     
which in turn suggests that also
\begin{equation}
m_{R^0}\equiv m^0(\tilde gg)<m^0\left[\tilde g(u\bar d)_8\right]=m_{R^+}
\end{equation}     
Fortunately we have also direct ``experimental'' evidence that the $R^+$ should
be heavier than the putative light $R^0$ by at least $m_\pi$. (see
eq.(\ref{53})). Indeed if the reverse, namely
\begin{equation}
m_{R^+}\le m_{R^0}+\pi^+,
\end{equation}     
is true, then we would have a new, (almost) stable, charged hadron of mass
$m_{R^\pm}\le1.6\pm0.2$ GeV.

Its production cross section -- particularly at large $p_T$ -- should
considerably exceed these of the deutron $d$ or antideutron $\bar d$. 
Such particles should have been detected for lifetimes in the range considered
\begin{equation}
3\times10^{-9}\le\tau_{R^+}\simeq\tau_{R^0}\le10^{-4}
\end{equation}     
Thus the $R^0$ is really conserved in hadronic collisions. In any reaction
initiated by $R^0$ the final state will also include an $R^0$. In particular,
any $R^+$ produced would decay via
\begin{equation}
R^+\to R^0\pi^+
\end{equation}
on hadronic time scales. 

One final comment concerns the small nuclear effects on $R^0$ nuclear
scattering at high and intermediate energies. These ``shadowing'' effects tend
to reduce the cross section of any projectile $x$ on a nucleus from the naive,
weak coupling value
\begin{equation}
\sigma^{(0)}(x,A)=A\;\sigma(x,p)
\end{equation}     
and correspondingly prolong the interaction mean free path from
\begin{equation}
l^{(0)}(x,A)=\frac{1}{n\;\sigma(x,p)}=\frac{1}{\rho Av\sigma(x,p)}
\end{equation}     
with $Av=$ Avogadro number $=6\times10^{23}$ and $\rho$ the material density so
that $n=\rho Av$ is a nucleons number density. The shadowing effects are
appreciable when
\begin{equation}
\sigma^0(x,A)\equiv A\;\sigma(x,p)\simeq\pi R^2(A,Z)=\pi A^{2/3}\;(1.2)^2\;
({\rm Fermi})^2
\end{equation} 
or
\begin{equation}
\sigma(x,p)\ge A^{-1/3}\cdot 45\;{\rm mb}\simeq(18-12)\;{\rm mb}
\end{equation}     
for $A=16-56$.

Thus we will use for iron longer interaction path than those implied by
eq.(\ref{82}), (\ref{80}):
\begin{equation}
\label{100}
l_{mfp(R^0-{\rm Iron})}=(10-30)\;{\rm cm}
\end{equation}     
with the lower (higher) value corresponds to $\sigma^I_{(R^0-N)}=40(7)$ mb with
substantial $\sim50\%$ (no) shadow corrections.

\section{The evolution of energetic $R^0$ particles in matter}
Energetic, $E_{\rm initial}(R^0)\ge30-40$ GeV, $R^0$ particles evolve as they
propagate in matter in three main stages corresponding -- in reverse order --
to the energy ranges of eqs.(\ref{22}),(\ref{23}) and (\ref{24}) above. We will
next try to follow the evolution in these stages -- utilizing the estimates of
the previous section.

{\bf Stage (a):}
Fast $R^0$-s are conserved through their hadronic cascade evolution -- if we
neglect the presumably small probability to generate further $R^0R^0$ pairs in
$R^0N$ collisions with c.m.s. energies
$W_{R^0N}\simeq(2E_{R^0}m_N)^{1/2}\le10$ GeV. This drastically simplifies their
evolution as compared with that of other long lived neutral particles such as
$K^0$-s, which are created in secondary hadronic collisions, and neutrons,
which, even neglecting secondary $\bar NN$ pair creation, can be
knocked out of nuclei. All these processes not-withstanding the hadronic showers
generated by $\bar qq$ jets (and also $\bar cc$, $\bar bb$ jets) essentially
quench at $l_{\rm que}\simeq10$ hadronic interaction lengths inside the
calorimeter or shielding materials.

We would like to argue that as they evolve through the energy range $40\;{\rm
GeV}\ge T_{R^0}\ge0.5$ GeV, the $R^0$-s will typically get further away from
the initial interaction point than $l_{\rm que}$, i.e. further than typical
hadronic showers. Several effects contribute to this:

\begin{description}
\item[(i)]
Charged particles continuously loose energy via ionization and stop at well
defined ranges -- even if they have no hadronic interactions. These losses
increase at low (sub GeV) energies: thus the ranges for 2 GeV $p$, $K^+$ and
$\pi^+$ in iron are:
\begin{equation}
R(p)=16\;{\rm cm}, \quad R(K^+)=30\;{\rm cm}, \quad  R(\pi^+)=60\;{\rm cm}
\end{equation}  
These ionization losses can indirectly contribute also to the slowing down of
$K^0$-s and neutrons. At $T_H\simeq1/2-2$ GeV hadron's kinetic energies, we have
appreciable cross sections for charge exchange reactions
\begin{equation}
K^0(\bar K^0)+N\to K^+(K^-)+N'
\end{equation}     
\begin{equation}
n+N\to p+N'
\end{equation}     
with $N$ referring to a nucleon in the nucleus. Also in inelastic collisions at
$T_H=2-40$ GeV the incoming $K^0$ or neutron often ``fragments'' into a leading
$K^\pm$ or proton. Thus roughly during half of the evolution of the hadronic
showers the nucleonic or ``kaonic'' component manifests as the charged
component of the isospin multiplets which slow down via ionization.

As emphasized towards the end of the previous section there are no analog
$R^0\to$stable $R^\pm$ conversion and there is no -- even indirect -- ionization
losses during the $R^0$ evolution \cite{cite}.

\item[(ii)]
The $R^0$ particles are likely to maintain higher fraction $x_F(R^0)$ of their 
initial momenta than nucleons or kaons.

\item[(iii)]
The $R^0$-nuclear inelastic cross sections in general and particularly in this
$40\;{\rm GeV}\ge T_{R^0}\ge1/2$ GeV range are likely to be smaller than $N-N$
and even $KN$ cross sections. 
\end{description}

To simplify the treatment of $R^0$ evolution in
stage (a) we will neglect the slow buildup of transverse momenta -- namely
momenta transverse to the initial entry direction of the $R^0$ particle into
the material. (After $n$ collisions we have
\begin{equation}
\overline{|p_T^{(n)}|^2}\simeq n\Delta_0^2
\end{equation}     
with $\Delta_0\simeq1/2$ GeV typical soft momentum transfer.) This leads then
to a one-dimensional evolution in momentum space [or more conveniently in
rapidity space $y_n=-\ln x_n$ with $x_n$ the fraction of the initial momentum
carried by the $R^0$ after $n$ collisions] and in coordinate space. The 
coordinate
is the distance $r$ (measured along the momentum of initial $R^0$) from the
entry point to the material, to the $R^0$ location. For energy independent
cross sections (particularly appropriate for $R^0N$ scattering which are
devoid of the varying Reggeon contribution) these $y$ and $r$ evolution
decouple. 
In a uniform medium, assumed for simplicity, the mean interaction length for
$R^{(0)}$ is then a constant, $l_0$. Let $g^{(n)}(r)$ be the
$r$ distribution after $n$ collisions. By definition
\begin{equation}
g^{(1)}(r)\simeq e^{-r/l_0}
\end{equation}     
We can readily derive the recursive relation
\begin{equation}
g^{(n+1)}(r)\simeq\int\limits_0^r dr'\;g^{(n)}(r')\;e^{-(r-r')/l_0}
\end{equation}     
and its solution
\begin{equation}
g^{(n)}(r)\simeq e^{-r/l_0}\;r^{n-1}
\end{equation}     
The average distance after $n$ steps is then
\begin{equation}
\overline{r^{(n)}}=nl_0=n\overline{r^{(1)}}
\end{equation}     
as expected. 

The rapidity space evolution will be completely analogous if we
assume that after a single collision we have rapidity distribution
$f^{(1)}(y)=e^{-\beta y}$ (corresponding to $x^\beta$ in $x$ space) so that
$\overline{y^{(1)}}=1/\beta$. 

Let us denote by $\lambda^{(a)}$ the energy-momentum average reduction factor
in stage (a)
\begin{equation}
\lambda^{(a)}\equiv\overline{x_F^{(R^0)}}
\end{equation}     
According to eq.(\ref{33}) we take $\lambda^{(a)}=0.65\pm0.15$. Starting with an
initial energy $E_0=T_0(=40\pm10$ GeV), we need then altogether $N^{(a)}$
collisions in stage (a) so as to degrade the kinetic energy to the required
$T_f(=1/2$ GeV) where
\begin{equation}
\label{109}
N^{(a)}=\frac{\ln(E_0/T_f)}{\ln(1/\lambda)}
\;(\simeq\ln(80\pm20)/\ln(1.6\pm0.3)\simeq7-18 )
\end{equation}     
[We will consistently use () brackets following equations for the actual
numerical estimates used in applying the equations.] The total penetration
distance traveled by the $R^{(0)}$ particle in stage (a) is then on average 
\begin{equation}
\label{110}
L^{(a)}\simeq N^{(a)}\;l^{(a)}_{\rm int}=\Bigl((12\pm5)l^{(a)}_{\rm int}
=(3\pm2.3)\;{\rm ``Iron}\;{\rm meter}\;{\rm equivalent''}\Bigr)
\end{equation}     
where we used eqs.(\ref{109}) and (\ref{100}). The very large spread in
(\ref{110}) represents the accumulated uncertainties in $\sigma^I(R^0N)$ or
$l^{(a)}_{\rm int}$ and in  the elasticities $\lambda^{(a})$ and finally
in the initial $R^0$ energy $E^{(0)}=40\pm10$ GeV. We believe that the values
$L^{(a)}_{\rm min}=0.7$ meter i.e. or $L^{(a)}_{\rm max}=5.3$ meter i.e. are
very unlikely. In particular to achieve $L_{\rm min}$ we need to assume
$\sigma^{\rm inelastic}_{R^0N}\simeq40$ mb (and $\bar x_F(R^0)\simeq1/2$) 
which are
too high (low) respectively and for $L_{\rm max}$ we need $\sigma^I\simeq7$ mb
(and $\bar x_F(R^0)\simeq 0.8$) which appear to be too low (high) respectively. 
We will adopt 
\begin{equation}
L^{(a)}=3\pm1.5\;{\rm meters}
\end{equation}     
The main conclusion which we would like to  emphasize is that
the putative $R^0$ particle travels beyond the full extent of
(e.m.+hadronic) showers due to normal (non $R^0$) hadronic jets. Indeed the
extent of the normal shower is affected also by ionization losses of the
leading particles to which the $R^0$-s are completely immune. Throughout most 
of the stage (a) the $R^0$-s move with $\beta_{R^0}\sim1$, hence its total
(Lab-time) duration is
\begin{equation}
t^{(a)}\simeq L^{(a)}/c\simeq10^{-8}(1{{+3}\atop{-0.4}})\;{\rm sec}
\end{equation}     
Because of time dilation effects the required proper time (and hence the lower
bound on $\tau_{R^0}$ required if indeed the $R^0$ particle is to complete this
stage) is lower than the above $t^{(a)}$. Specifically we find that
\begin{eqnarray}
\tau^{(a)}=\sum\limits_{i=0}^{N^{(a)}-1}\tau_i
=\frac{l_0}{c}\sum\limits_0^{N^{(a)}-1}(\lambda^{(a)})^i
=\frac{l_0}{c}\Bigl[\frac{1-(\lambda^a)^{N^{(a)}}}{1-\lambda^a}\Bigr]
\nonumber\\
\simeq(2-5)\frac{l_0}{c}\simeq(0.6-5)\times10^{-9}\;{\rm sec}
\end{eqnarray}     
where in the last equation, $\tau_i$, the proper time lapse between the 
$N_a-i$th and $N_a-i+1$th collision, is shortened relative to $l_0/c$ by
$1/\gamma^{(i)}\simeq(\lambda_a)^i$ and we neglected
$(\lambda_a)^{N_a}=\frac{T_f}{E_0}\simeq\frac{1}{80}$. Thus the modest 
requirement
\begin{equation}
\tau_{R^0}\ge5\times10^{-9}\;{\rm sec}
\end{equation} 
suffices to ensure that the initial $R^0$ particles survive stage (a).

{\bf Stage (b):} Intermediate energies. 

Using eq.(\ref{43}) for the average fraction of kinetic energy
$\lambda^{(b)}(=0.62\pm0.1)$ retained in quasi-elastic $R^0$-nucleus collisions
in stage (b) we need on average
\begin{equation}
\label{115}
N^{(b)}=\frac{\ln\left[T_{\rm initial}^{(b)}/T_{\rm final}^{(b)}\right]}
{\ln(1/\lambda^{(b)})}(=\frac{ln50}{\ln1.6}\simeq8)
\end{equation}      
collisions to complete this stage. We used in (\ref{115}) 
$T_{\rm initial}^{(b)}=0.5$ GeV, $T_{\rm final}^{(b)}=0.01$ GeV for the kinetic
energies characterizing the entry to and exit from ``stage (b)''. The
kinematics of the elastic $R^0$-nucleon collision implies that for
$\frac{d\sigma}{d\Omega^*}(R^0)\simeq$constant (i.e. constant differential
cross section in the cms frame) we have in the lab frame for $u=m_{R^0}/m_N$
\cite{landau}:
\begin{equation}
f_u(\cos\theta)\equiv
\frac{d\sigma}{d\Omega}|_{R^0}
=\frac{1+u^2\cos2\theta}{(1-u^2\sin^2\theta)^{1/2}}
(\simeq\frac{1+(2.5\pm0.6)\cos2\theta}{(1-(2.5\pm0.6)\sin^2\theta)^{1/2}})
\end{equation}    
Let the initial $R^0$ direction be $\hat n_0$, and $\hat n_k$ be the direction
after $k$ successive scatterings. In $f_u(\cos\theta)$ the $\cos\theta$
argument refers to $\hat n_k\cdot\hat n_{k-1}$. The distribution after $k$
steps will be
\begin{eqnarray}
f^{(k)}(\cos\theta)\equiv f^{(k)}(\hat n_k\cdot\hat n_0)
=\int\;d\hat n_1\;d\hat n_2\dots d\hat n_k
\nonumber\\
\times f_u(\hat n_0\cdot\hat n_1)\; f_u(\hat n_1\cdot\hat n_2)
\dots f_u(\hat n_{k-1}\cdot\hat n_k)
\end{eqnarray}    
Using the completeness relation
\begin{equation}
\frac{1}{4\pi}\int\;d\hat n'\; P_l(\hat n\cdot\hat n')\; 
P_{l'}(\hat n'\cdot\hat m)=\frac{1}{2l+1}\delta(l,l')\;P_l(\hat n\cdot\hat m)
\end{equation}    
we find that if
\begin{equation}
f_u(\cos\theta)=\sum\limits_{l=0}^\infty\;f_l(u)\;(2l+1)\;P_l(\cos\theta)
\end{equation}    
then
\begin{equation}
f^{(k)}(\cos\theta)=\sum\limits_{l=0}^\infty\left[f_l(u)\right]^k\;
(2l+1)\;P_l(\cos\theta)
\end{equation}    
We will be particularly interested in the net extra displacement of $R^0$ --
namely the extra displacement along the initial $R^0$ direction ($\hat n_0$)
accumulated during stage (b): $\Delta L^{(b)}=l_0\sum\limits_{k=1}^{N^{(b)}}\;
\cos(\widehat{n_k\;n_0})$. On average it will be 
\begin{equation}
\overline{\Delta L^{(b)}}=l_0\sum\limits_{k=1}^{N^{(b)}}\;
\overline{(\hat n_k\cdot\hat n_0)}=l_0\sum\limits_{k=1}^{N^{(b)}}\;(f_1)^k
=\frac{l_0\;f_1}{1-f_1}(1-f_1^N)
\end{equation}    
with
\begin{equation}
f_1(u)=\int\;d(\cos\theta)\;f_u(\cos\theta)\;P_1(\cos\theta)
=C\int\limits_a^1\;dx\;x\;\frac{1+u^2(2x^2-1)}
{\left[(1-u^2)+u^2x^2\right]^{1/2}}
\end{equation}    
where
$$a\equiv\frac{(-1+u^2)^{-1/2}}{u}$$
and $C$ is a normalization constant

\begin{equation}
C^{-1}=\int\limits_a^1\;dx\;\frac{1+u^2(2x^2-1)}
{\left[(1-u^2)+u^2x^2\right]^{1/2}}
\end{equation}
The transverse displacement is given by a ``random walk''
\begin{equation}
(\Delta_T^{(b})^2=(l_0^{(b})^2\sum\limits^{N^{(b)}}\sin^2(\widehat{n_k\;n_0})
(\le\frac{2}{3}N^{(b)}l_0^2)
\end{equation}   
using the above with $l_0^{(a)}\simeq l_0^{(b)}\simeq10-30$ cm, $u=1.6\pm0.3$,
$N^{(b)}=8\stackrel{+6}{-3}$ we estimated:
\begin{equation}
\Delta L^{(b)}\simeq0.4-1.3 \;{\rm meters}\;{\rm i.e.}
\end{equation}   
\begin{equation}
\Delta_T^{(b)}\simeq0.5\pm0.25\;{\rm meters}
\end{equation}   
Finally the $R^0$ motion during most of stage (b) non-relativistic. The 
velocity after the $n$-th collision is therefore:
\begin{equation}
\beta_n=\beta_0(\sqrt{\lambda^{(b)}})^n
\end{equation}   
The total time ($\simeq$proper time) consumed in this stage is
\begin{equation}
\Delta t^{(b)}=\Delta\tau^{(b)}=\sum\Delta t_n^{(b)}=\frac{l_0}{\beta_0 c}\;
\frac{1-(\lambda_a)^{(\frac{N^{(b)}}{2}+\frac{1}{2})}}{1-(\lambda^{(b)})^{1/2}}
(\simeq1.5-5\times10^{-8}\;{\rm sec})
\end{equation}   
with $\beta_0\simeq0.8$ the initial velocity corresponding to $T_{R^0}=0.5$
GeV.

Thus if
\begin{equation}
\tau_{R^0}\ge\tau^{(a)}+\Delta\tau^{(b)}\simeq(2.5{{+6}\atop{-0.5}})
\times10^{-8}\;{\rm sec}
\end{equation}   
the $R^0$ particle would, on average, live through the end of stage (b). Its
distance from the entry point is then:
\begin{equation}
\label{129}
L_{ab}=L^{(a)}+\Delta L^{(b)}\simeq4\pm1.6\;{\rm meters}
\end{equation}   
with a transverse spread of
\begin{equation}
\label{130}
\Delta_T=0.5\pm0.25\;{\rm meters}
\end{equation}   

{\bf Stage (c):}
Our discussion in sec.4 suggests that once $T_{R^0}\le8$ MeV (or the $R^0$ 
velocity satisfies
$\beta\le\beta_0=0.1$) the $R^0$ particles will be quickly captured into
$R^0-(A,Z)$ bound states. This would be effectively ``fix'' the $R^0$ and
prevent further diffusion away from the $R^0$ location it reaches by the end
of stage (b) at the ($L,\Delta_T$) values quoted above (eq.(\ref{129}),
(\ref{130})).

Even if the $R^0$ particles stay unbound for the duration of its lifetime 
$\tau_{R^0}$, the extra
diffusion distance $\sqrt{N^{(c)}}l_0$ grows very slowly with $\tau_{R^0}$ like
$\sqrt{\ln\tau_{R^0}}$. From eq.(\ref{74}) above we find that the time
intervals between successive $R^0$-nuclear collisions are:
\begin{equation}
\Delta t^{(n)}=\Delta t^0\left(1+\frac{u}{A}\right)^n\equiv\frac{l_0}{\beta_0c}
\;\left(1+\frac{u}{A}\right)^n
\end{equation}    
Thus all $N^{(c)}$ collisions of stage (c) will be completed after a time
\begin{equation}
t^{(c)}\simeq\tau_{R^0}=\sum\limits^{N^{(c)}}\Delta t^{(n)}
\Bigl\{(1+\frac{u}{A})^{N^{(c)}+1}-1\Bigr\}\frac{A}{u}=
\Delta t^{(0)}\frac{A}{u}e^{\frac{u}{A}N_c}
\end{equation}   
where $N^{(c)}\gg1$ is implicit. Specifically we find for $\tau_{R^0}=10^{-4}$ 
sec
\begin{equation}
\label{133}
N^{(c)}=\frac{A}{u}\ln\Bigl(\frac{\tau_{R^0}}{\Delta t^0}\frac{u}{A}\Bigr)
(\simeq 80\;{\rm for}\;\;A=56\;\;l_0=15\;{\rm cm}\;\;u=m_{R^0}/m_N\simeq1.6)
\end{equation}   
where $l_0$ was estimated via the geometric $R^0-(A,Z)$ cross section
$\sigma=\pi R^2(A)=3.6\times10^{-25}\;{\rm cm}^2$ to be
\begin{equation}
l_0\simeq15\;{\rm cm}\;\;{\rm in}\;{\rm iron}
\end{equation}   
For $\tau_{R^0}=10^{-6}$ sec we find instead of eq.(\ref{133})
\begin{equation}
N^{(c)}=40\;({\rm for}\;\;A=56,\;\;l_0=15\;{\rm cm},\;\;
\Delta t_0=5\times10^{-9}\;{\rm sec})
\end{equation}   
The maximum total extra diffusion distance added in stage (c) is therefore
\begin{equation}
|\Delta \vec R^{(c)}|=\sqrt{N^{(c)}}l_0=
\left\{ {1.4\;{\rm meters},\qquad\tau_{R^0}=10^{-4}\;{\rm sec}}
 \atop{ 0.9\;{\rm meters},\qquad\tau_{R^0}=10^{-6}\;{\rm sec}} \right\}
\end{equation}   
In passing we note that once $R^0$ (or a neutron) achieves a final thermal
velocity ($\simeq2$ km/sec, or $\beta_f=0.6\times10^{-5}$) one reverts again to 
ordinary diffusion with $\Delta R\simeq\sqrt{\tau}$. However even after 100
elastic collisions with iron nuclei a $R^0$ particle with starting velocity
$\beta_0=0.1$ will on average go down only to $\beta_f=0.005$.

A key observation is that if the $R^0$ decays at any time during stage (c) we
expect its final velocity to be
\begin{equation}
\beta_f\le0.1=\beta_0
\end{equation}

Furtuitisely this holds also if the $R^0$ is bound to the nucleus with binding
energy $|\epsilon|\simeq8$ MeV. The Fermi momentum of the bound $R^0$
($\simeq\sqrt{2|\epsilon|m_{R^0}}\simeq0.16$ GeV) corresponds to 
\begin{equation}
\beta_f\simeq0.1
\end{equation}
Consequently the Lab energy of the $\pi^0$ from the decay
$R^0\to\pi^0\tilde\gamma$ will be rather monochromatic. Specifically
\begin{equation}
E_{\pi^0}=(1+\beta_f\cos\theta)E^{(0)}_{\pi^0}\equiv
(1+\beta_f\cos\theta)\frac{m^2_{R^0}-m^2_{\tilde\gamma}+m^2_{\pi^0}}{2m_{R^0}}
\end{equation}   
satisfies (on average) 
$|E_{\pi^0}(\beta,\cos\theta)-E_{\pi^0}^{(0)}|/E^{(0)}_{\pi^0}
\le\frac{\beta_f}{2}=0.05$.

We conclude this section by a short discussion of the likely location of this
$R^0$ decay event, $\vec r$ (decay), measured relative to the intersection
point. We will consider first the longitudinal displacement $r_L$ along the
initial $R^0$ direction, $\hat n_0$. It is accumulated only in stages (a) and
(b) and assuming $\tau_{R^0}\ge10^{-8}$ sec so that stage (b) was completed
we have from eq.(\ref{129})
\begin{equation}
r_L\simeq L_{(a)+(b)}\equiv L_{ab}\simeq0.4\pm1.6\;{\rm meters}
\end{equation}   
The total displacement has also random diffusive ``components'' from stage (b)
and (c). Thus finally the total displacement is
\begin{equation}
\label{140}
\langle\vec r^2\rangle^{1/2}\simeq r
\simeq\sqrt{(L_{ab})^2+(\Delta_B^T)^2+(\vec R_c)^2}(\simeq4.5\pm2\;{\rm meters})
\end{equation}   
And its average direction relative to $\hat n_0$ is 
\begin{equation}
\label{141}
\hat n_0\cdot\hat r\simeq r_L/r\simeq0.8
\end{equation}   

The above estimates referred to iron, and also neglected the freely traveled
distance to the calorimeters and hence $r$ is underestimated. We note
that lighter (concrete rock+earth materials) may actually expedite the slowing 
down of $R^0$ particles in stage (c) because the fraction of energy loss is 
$\simeq\frac{1}{A}$.

\section{The proposed $R^0$ search strategie}
The search strategy proposed here address a wide range of putative $R^0$ 
parameters. Specifically we will assume $\tau_{R^0}\ge3\times10^{-8}$ sec,
$\sigma_{R^0N}<\sigma_{NN}$ for both elastic/inelastic cross sections and
$\bar x_F(R^0)\ge\bar x_F(p)=0.5$ for the momentum fraction retained by the
leading $R^0$ particle in inelastic collisions. The search is optimized if
$\sigma_{R^0N}\simeq\frac{1}{2}\sigma_{NN}$, $\bar x_F(R^0)\simeq0.7$ and
particularly when $Br(\pi^0)$, the branching ratio for the
$R^0\to\pi^0\tilde\gamma$ decay mode is large, say $Br(\pi^0)\ge20\%$.

Let us assume that a detector ``D'' of volume $V_D$ has been installed
underground at, say, the FNAL collider (or at $e^+e^-$
colliders) at a distance $R$(=5$\pm$2 meters) from the intersection point and
with $\vec R$ perpendicular to the beam direction. This detector searches 
$R^0\to\pi^0\tilde\gamma,\;\eta^0\tilde\gamma$ and possibly also 
$R^0\to\pi^+\pi^-\tilde\gamma$, $R^0\to\pi^+\pi^-\pi^0\tilde\gamma$ decays.
The signature distinguishing such decays from background events could be
an almost monochromatic $\pi^0$ with $E_{\pi^0}\ge0.6$ GeV. Alternatively final
states including several pions (and possibly some nucleons knocked out when the
decay $R^0\to\pi^+\pi^-\pi^0\tilde\gamma$ occurres inside the nucleus, once
the ($R^0-A,Z$) bound state formed) should have a relatively isotropic
distribution with limited missing momenta $p_{\tilde\gamma}\le
m_{R^0}/4\simeq0.4$ GeV. Our arguments for a longer migration length for
$R^0$-s as compared with the extent of hadronic showers suggest a reduction of
the neutron background at the location of the detector if it as placed at a
distance $r_0$ given by eq.(\ref{140}) above.

The collision of such neutrons with nuclei in our detector can produce $\pi^0$,
$\pi^+\pi^-$, etc. final states and constitute a background to the
$R^0$ decays. However only neutrons with  kinetic energies $T_{R^0}>$GeV 
($T_{R^0}>2$ GeV) are likely to produce $\pi^0$ ($\pi^+\pi^-\pi^0$) in $R^0N$
collisions. These background events will have achromatic $\pi^0$ or apparent
missing momenta -- often pointing in the direction of $\hat r$ i.e. towards 
the decay vertex. 

One can further reduce the energetic neutrons background in our detector by
excluding time ``windows'' $\Delta t$, say of magnitude $\Delta
t\simeq3\times10^{-8}$ sec following each collision event or, if the rep cycle
of the collider is short (as in the upgraded FNAL Tevatron), the first 
$\Delta t\simeq3\times10^{-8}$ sec following relevant events with jets. [Only
1-2 GeV neutrons from such events emitted in the direction of the jets are 
likely to make it to the required
transverse distance from the colliding beam. In typical minimal bias events
the neutron along with most other particles produced in the beam's jets
are forward backward peaked]. The total distance traveled by relativistic 
hadrons in
this $\Delta t$ is 9 meters, and most of the energetic neutrons are likely to 
have interacted and slowed down by then. Yet the slowly diffusing $R^0$ may 
well have not decayed if
$\tau_{R^0}\ge3\times10^{-8}$ sec. This negative time correlation will also
reduce background from energetic muons emerging from the primary collision 
which interact in the
vicinity of our detector and produce secondary neutrons. Such muons, cosmic ray
muons, and charged particles background in general, can be further reduced by
an anti-coincidence network surrounding our detector at a distance of $\simeq$1
meter from its boundaries. Also the very strong, slow or even thermal,
ambient neutron background -- causing many low energy depositions, via
nuclear captures -- can be reduced by using a proper neutron absorber shielding.

Finally we note that having over most of the distance $\vec r$ between the 
intersection
point and the detector iron magnetized in a direction perpendicular to $\vec r$
could also drastically reduce the neutron
background by having the intermediate proton curve in the magnetic field
\cite{cite}.

All this notwithstanding we cannot hope to conclusively ``pin down'' $R^0$
particles by looking at their decays only. The basic reason is that,
even if the detector is placed at an optimal distance $r$ (see eq.(\ref{140}))
(which is {\it a priori} known only within $\pm$60\%!), and the $R^0$ particles
are isotropically distributed, only a small fraction 
\begin{equation}
f=\frac{V_D\;Br(R^0\to\pi^0)}{\frac{4\pi}{3}r^3}\simeq[V_D/m^3]\cdot
(3\times10^{-5}-2\times10^{-3})
\end{equation}       
of the $R^0$-s will decay via the $\pi^0\tilde\gamma$ mode, inside the
detector. [In the above estimate we used $r=5\pm3$ meters, $Br=0.1\pm0.05$].

To pick out the genuine $R^0$ decays we correlate events in the detector with
primary $\bar pp$ collision both temporally and directionally.
Specifically we will focus on primary $\bar pp$ collisions with $\tilde g$ 
jets that
could give rise to a sufficiently high energy $R^0$ that in turn could make it
to our detector. 

Thus we will consider those ``special'' $\bar pp$ events with one (or two)
jet(s) of transverse energy $E_T\ge E_T^{(0)}(\simeq10$ GeV), and with no 
energetic
lepton(s) [to avoid b,c jets] in the jets(s). If a light gluino exists these
would be gluino jet(s) with an appreciable $\simeq$20\% probability. Beam
crossing and one $\bar pp$ collision on average happens every 
$$\delta_{t_H}=3\times10^{-6}(10^{-7})\;{\rm sec}$$
in the old (upgraded) versions. However only one in $N_H\simeq10^6$ collisions
is ``special'' in the above sense. ($N_H$ is the ratio of total $\bar pp$
cross section to a perturbative estimate for the cross section of the 
gluino-jet production).

Hence the ``special'' collisions will be spaced, on average, by $\Delta
t_H=N_H\delta t_H\simeq3-(0.1)$ sec. If the $\Delta E\ge0.6$ GeV energy 
deposition
in our detector is due to a decay of an $R^0$ from a gluino jet in a particular
``special'' event then we expect that:

\begin{description}
\item[(i)]
The decay should happen on average at time $\simeq\tau_{R^0}$ after the primary
special collision.

\item[(ii)]
The jet (or at least one jet) in the primary event should point in the
direction of our detector within a $\Delta\theta^0$ uncertainty with
$\cos(\Delta\theta^0)$ given by eq.(\ref{141})

\end{description}

Jointly these two requirements will reduce the background by
\begin{equation}
(\tau_{R^0}/\Delta t_H)\;(\Delta\Omega/4\pi)\;
(\simeq\frac{3\times10^{-8}-10^{-4}}{3-0.1})\;(\frac{1}{20}-\frac{1}{10})
\simeq(5\times10^{-10}-10^{-4})
\end{equation}
where 
\begin{equation}
\Delta\Omega=2\pi[1-\cos(\Delta\theta^0)]=2\pi(1-\hat n_0\cdot\hat r_0)
=2\pi(\frac{1}{10}-\frac{1}{15})
\end{equation}
This background reduction may suffice for facilitating an $R^0$ research.

Finally if the hadronic showers in the special events which do correlate with 
decays in the detector D extend further into the calorimeters than the average 
hadronic
(non,b,c) jets, as we would expect on the basis of the $R^0$ evolution, we
will have an additional beautiful confirmation!

Needless to say all this very qualitative arguments need verification via
detailed Monte-Carlo studies with $\tilde g$ jets and evolving $R^0$ particles.

The whole analysis can be repeated for $e^+e^-$ collisions at the $Z^0$
resonance.
We need now to specifically look for 4 jet events: $e^+e^-\to q\bar q\tilde
g\bar{\tilde g}$ which may be relatively rare occuring in only few percent of 
the
events. However we gain by having a much reduced hadronic background and a
larger $\delta t$. [The very large e.m. radiative background can be effectively 
avoided by thick lead shieldings around the detector].

In summary we have suggest a novel, and in our view optimized, alternative $R^0$
search
method. Instead of searching for $R^0$ decays in fixed target neutral beam
where the putative $R^0$-s are at a considerable disadvantage, we suggest using
colliders and jets at $90^o$ where $R^0$ could be relatively copious. Also by
moderating first the $R^0$ particles a very wide range of $R^0$ lifetimes
becomes accessible.

While many of the basic $R^0$ features appear in the papers by G.Farrar we 
included
them for completeness. The discussion of sections 4-5 has new elements --
and in particular the possibility of having longer attenuation path for $R^0$ in
matter as compared with neutrons is discussed at length.

Negative results in the experiments suggested cannot unequivocally exclude the 
eight gluino hypothesis. Though we argued for $R^0N$ 
cross sections smaller than $N-N$ cross sections the reverse may be true and 
the $R^0$ will then migrate less than our estimates making it more difficult 
to separate $R^0$ decays from the ``normal''hadronic activity. Also for 
$\tau_{R^0}\simeq3\times10^{-9}-3\times10^{-8}$ sec direct searches for decay 
in flight may be more efficient. 

Eventually many different lines of research will converge to yield a definitive 
verdict on the light gluino hypothesis. At present however the $R^0$ may exist. 
It will be a great pity if it will be missed due to lack of enthusiasm and 
possibly for not trying the right optimal approach to discovering it.

\section{Acknoledgements}

I am indebted to J.Barnet, Z.Bern, D.Gerdes, A.Grant, S.Ino, A.Morgan, 
E.Pevsner, C.Quigg, J.Sandweis, M.Voloshin and C.Whitten 
for many useful discussions and to J.Bagger for explaining the 
light gluino scenario. 

I am particularly thankfull to L.Madansky for crucial insight and suggestions.


\begin{thebibliography}{99}
\bibitem{cite}
An even more drastic slowing down and partial screening of the medium energy
neutrons via protons that they convert into and regenerate from, can be
achieved if the iron shielding is strongly magnetized.
\end{thebibliography}
\end{document}